\documentclass[twocolumn,showpacs,preprintnumbers,amsmath,amssymb,floatfix]{revtex4}

\usepackage{graphicx}
\usepackage{dcolumn}
\usepackage{bm}

\begin{document}

\title{Novel insights into charge and spin pairing instabilities
in nanoclusters}

\author{A.~N.~Kocharian$^{1}$, G.~W.~Fernando$^{2}$,  K.~Palandage$^2$ and J.~W.~Davenport$^3$}
\affiliation{$^1$Physics Dept., California State University, Los
Angeles, CA 90032}
\affiliation{$^2$U-46, Physics Dept.,
University of Connecticut, Storrs, CT 06269}
\affiliation{$^3$Computational Science Center and Center for
Functional Nanomaterials, Brookhaven National Laboratory, Upton,
NY 11973}

\begin{abstract}
Electron pairing and ferromagnetism in various cluster geometries
are studied with emphasis on tetrahedron and square pyramid under
variation of interaction strength, electron doping and
temperature. These exact calculations of charge and spin
collective excitations and pseudogaps yield intriguing insights
into level crossing degeneracies, phase separation and
condensation. Criteria for spin-charge separation and
reconciliation driven by interaction strength, next nearest
coupling and temperature are found. Phase diagrams resemble a
number of inhomogeneous, coherent and incoherent nanoscale phases
seen recently in high T$_c$ cuprates, manganites and CMR
nanomaterials.
\end{abstract}

\maketitle
\section{Introduction}
Systems with correlated electrons display a rich variety of
physical phenomena {{and properties}}: different types of magnetic
ordering, (high-Tc) superconductivity, ferroelectricity,
spin-charge separation, formation of spatial inhomogeneities
~\cite{Davis1,Valla,Bodi,hashini,Yazdani,deHeer0} (phase
separation, stripes, local gap and incoherent pairing, charge and
spin pseudogaps). The realization of these properties in clusters
and bulk depends on interaction strength $U$, doping, temperature,
the detailed type of crystal lattice  and sign of coupling
$t$~\cite{Nagaoka}. Studies of perplexing physics of electron
behavior in non-bipartite lattices encounter enormous
difficulties. Exact solutions at finite temperatures exist only in
a very few cases~\cite{Shiba,Falicov,Schumann1}; perturbation theory
is usually inadequate while numerical methods have serious
limitations, such as in the Quantum Monte-Carlo method and its
notorious sign problem. On the contrary, one can get important
insights from the exact solutions for small clusters
(``molecules"). For example, squares or cubes are the building
blocks, or prototypes, of solids with bipartite lattices, whereas
triangles, tetrahedrons, octahedrons without electron-hole
symmetry may be regarded as primitive units of typical frustrated
systems (triangular, pyrochlore, perovskite). Exact studies of
various cluster topologies can thus be very useful for
understanding nanoparticles and respective bulk systems. One can
take a further step and consider
 an inhomogeneous
 bulk system as a collection of
many such decoupled clusters, which do not interact directly, but
form a system in thermodynamic
equilibrium~\cite{Schumann1,JMMM,PRB}. Thus we consider a
collection of such ``molecules", not at fixed average number of
electrons per each cluster, but as a grand canonical ensemble, for
fixed chemical potential $\mu$. The electrons can be splintered
apart by spin-charge separation due to level crossings driven by
$U$ or temperature, so that the collective excitations of electron
charge and spin of different symmetries can become quite
independent and
propagate incoherently. We have found that local charge and spin
density of states or corresponding susceptibilities
can have different pseudogaps which is a sign of spin-charge
separation. For large $U$ near half filling, holes prefer to be
localized on separate clusters having Mott-Hubbard (MH) like
charge pseudogaps~\cite{PRB} and Nagaoka ferromagnetism (FM)
~\cite{Nagaoka}; otherwise, spin density waves or spin liquids may
be formed. At moderate $U$, this approach leads to reconciliation
of charge and spin degrees with redistribution of charge carriers
or holes between square clusters. The latter, if present, can
signal a tendency toward phase separation, or, if clusters
``prefer" to have two holes, it can be taken as a signature of
pairing~\cite{pairing}. This, in turn, could imply imposed
opposite spin pairing followed by condensation of
charge and spin degrees into a BCS-like coherent state. Although
this approach for large systems is only approximate, it
nonetheless gives very important clues for understanding large
systems whenever correlations are local. We have developed this
approach in Refs.~\cite{JMMM,PRB,PLA,PRL,COMP} and successfully
applied to typical unfrustrated (linked squares) clusters. Our
results
are directly applicable to
nanosystems which usually contain many clusters, rather well
separated and {{isolated}} from each other but nevertheless being
in thermodynamic equilibrium with the possibility of having
inhomogeneities
%
for
different
number
of electrons per cluster. Interestingly,
an ensemble of square clusters displays ``checkerboard" patterns,
nanophase inhomogeneity, incoherent pairing and nucleation of
pseudogaps~\cite{Davis1,Valla,Bodi,hashini,Yazdani}. The purpose
of this work is to further conduct similar extensive
investigations in frustrated systems~\cite{Serg}, exemplified by
4-site tetrahedrons and 5-site square pyramids. As we shall see,
certain features in
various topologies are quite different and these predictions could
be exploited in
the nanoscience frontier by synthesizing clusters or
nanomaterials with unique properties~\cite{TM}.
\begin{figure} 
\begin{center}
\includegraphics*[width=20pc]{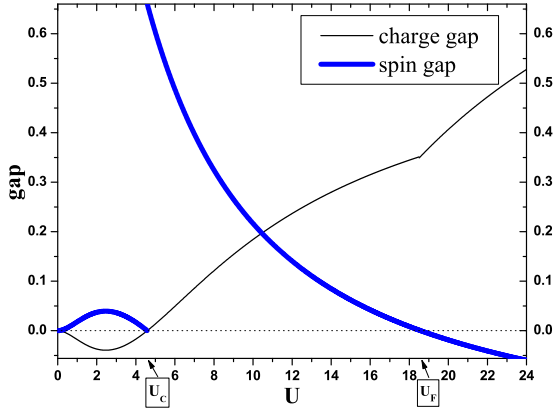}
\hfill
\end{center}
\caption {Charge $\Delta^{c}$ and spin $\Delta^{s}$ gaps versus
$U$ in an ensemble of squares at $\left\langle
N\right\rangle\approx 3$ and $T=0$. Phase A: Charge and spin
pairing gaps of equal amplitude at $U\leq U_c$ describe bose
condensation of electrons similar to BCS-like coherent pairing
with a single energy gap. Phase B: Mott-Hubbard like insulator at
$U_c<U<U_F$ leads to $S={1\over 2}$ spin liquid behavior. Phase C:
Parallel (triplet) spin pairing ($\Delta^s<0$) at $U>U_F$ displays
$S={3\over 2}$ saturated ferromagnetism (see Sec.~\ref{squares}).}
\label{fig:gap-4-site}
\end{figure}
\begin{figure} 
\begin{center}
\includegraphics*[width=18pc]{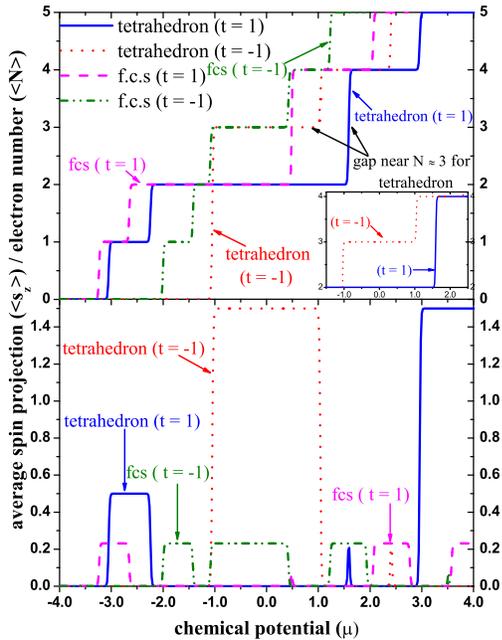}

\end{center}
\caption {$\left\langle N\right\rangle $ and $\left\langle
s^z\right\rangle $ versus $\mu$ in grand canonical ensemble of
tetrahedrons and fcs at $U=4.0$, $T=0.01$ and $h=0.1$.
Mott-Hubbard like ferromagnetism for $\left\langle
s_z\right\rangle={3\over 2}$ at $\left\langle N\right\rangle=3$ in
tetrahedron occurs for $t=-1$, while absence of charge pseudogap
near $\left\langle N\right\rangle\approx 3$ metallic state with
spin rigidity manifests level crossing degeneracy related to
pairing (see inset).} \label{fig:num_mag}
\end{figure}
\begin{figure} 
\begin{center}
\includegraphics*[width=18pc]{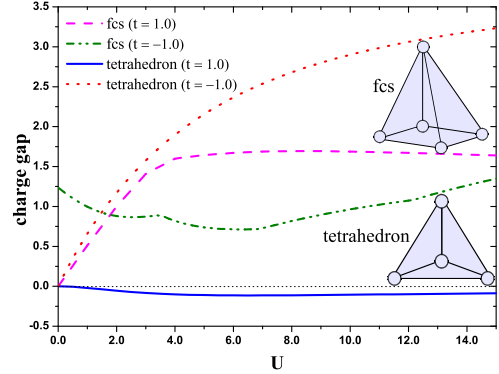}
 \hfill
\end{center}
\caption {$\Delta^{c}$ versus $U$ for one hole off half filling in
tetrahedrons and fcs at $T=0$. In tetrahedron, $\Delta^{c}<0$ at
$t=1$ implies phase separation and coherent pairing with
$\Delta^s\equiv \Delta^P$, while $\Delta^{c}>0$ for $S={3\over 2}$
at $t=-1$ leads to a ferromagnetic insulator ($\Delta^s<0$) for
all $U$. In fcs, $\Delta^{c}>0$ at $\left\langle
N\right\rangle\approx 4$ for $t={\pm}1$ describes Mott-Hubbard
like antiferromagnetism ($\Delta^{s}>0$).}\label{fig:gap}
\end{figure}
\begin{figure} 
\begin{center}
\includegraphics*[width=18pc]{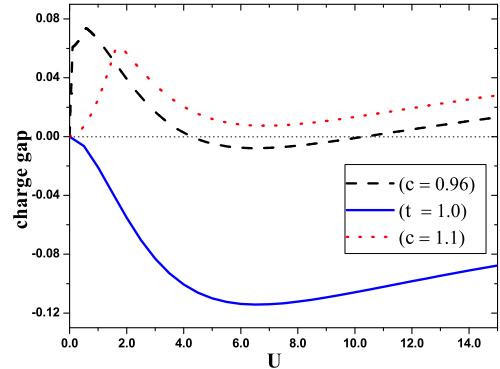}
\hfill
\end{center}
\caption {$\Delta^c$ versus $U$ at $\left\langle N\right\rangle=3$
and $T=0.01$ in tetrahedron ($t=1$) and deformed tetrahedral
clusters ($c=0.96$ and $1.1$). Charge and spin pairing gaps of
equal amplitude at $t=1$ imply coherent pairing, while
$\Delta^c>0$ and $\Delta^{s}<0$ at $c=1.1$ correspond to an
unsaturated ferromagnetic insulator for $S={1\over 2}$. Coherent
pairing is retained in a narrow range near $c\approx
1$.}\label{fig:pyramid}
\end{figure}
\begin{figure} 
\begin{center}
\includegraphics*[width=20pc]{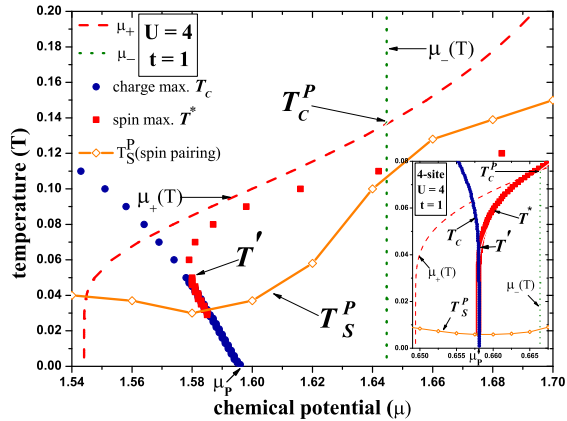}
 \hfill
\end{center}
\caption {The  $T$-$\mu$ phase diagram of tetrahedrons without
electron-hole symmetry at optimally doped $\left\langle
N\right\rangle\approx 3$ regime near $\mu_P=1.593$ at $U=4$ and
$t=1$ illustrates the condensation of electron charge and onset of
phase separation for charge degrees below $T_c^P$. The incoherent
phase of preformed
pairs with unpaired opposite spins exists
above ${T_s}^P$. Below $T_s^P$, the paired spin and charge coexist
in a coherent pairing phase. The charge
and spin
susceptibility peaks, denoted by $T^*$ and $T_c$, define
pseudogaps calculated in the grand canonical ensemble, while
$\mu_+(T)$ and $\mu_-(T)$ are evaluated in canonical ensemble.
Charge and spin peaks reconcile at $ T^\prime\geq T\geq T_{s}^P$,
while $\chi^c$ peak below $T_s^P$ signifies metallic (charge)
liquid (see inset for square cluster and
Ref.~\cite{PRL}).}\label{fig:ph4_6}
\end{figure}
\begin{table}
\caption{Ground state (GS) in various cluster geometries for one
hole off half filling at large $U=900$ and $t={\pm}1$ having
saturated ferromagnetism (SF), unsaturated ferromagnetism (UF),
antiferromagnetism (AF) or coherent pairing (CP). }\label{Nagaoka}
\begin{center}
\begin{tabular}{|c|c|c|c|c|c|c|c|}
\hline cluster type & N$_a$ & N & S&$\Delta^s$  & $\Delta^c$ &t & GS \\
\hline triangle & 3 & 2 & 1 &-0.998 & 3.011 &-1 & SF\\
\hline tetrahedron/nnn square & 4 & 3 &1.5 & -0.997 & 3.987&-1 & SF\\
\hline square pyramid/fcs & 5 & 4 & 2 & -0.417  & 2.596& -1  & SF \\
\hline  square & 4 & 3 & 1.5 &-0.262& 1.15 & ${\pm}1$& SF \\
\hline triangle & 3 & 2 & 0 & 1.008 & 1.993& 1 & AF\\
\hline tetrahedron/nnn square& 4 & 3 & 0 & 0.002 & -0.002  &1 & CP\\
\hline square pyramid/fcs & 5 & 4 & 1 & -0.115  & 1.543& 1  & UF \\
\hline
\end{tabular}
\end{center}
\end{table}

 \section{Model and formalism}
Exact diagonalization of the Hubbard model (HM)
\begin{eqnarray}
H=-t\sum\limits_{\left\langle ij\right\rangle
\sigma}c^{+}_{i\sigma}~c_{j\sigma}+U \sum\limits_{i} n_{i\uparrow}
n_{i\downarrow},\label{2-site1}
\end{eqnarray}
and quantum statistical calculations of charge ${\chi_c}$ and spin
$\chi_s$ susceptibilities, i.e., {\it fluctuations}, in
a grand canonical ensemble
and pseudogaps $\Delta^c(T)=E(N+1)+E(N-1)-2E(N)$ and
$\Delta^s(T)=E(S+1)-E(S)$ for canonical energies $E(N)$ and $E(S)$
($N$ and $S$ being the total number of electrons and spin
respectively)
yield valuable insights into
quantum critical points and various phase transitions as shown in
Ref.~\cite{PRB}. By monitoring the peaks in $\chi^c$ and $\chi^s$
one can identify charge/spin pseudogaps and relevant crossover
temperatures; nodes, sign and amplitude of pseudogaps
determine energy level crossings, phase separation, electron
pairing ranges, spin-charge separation and reconciliation regions.

\section{Results}
\subsection{Bipartite clusters}~\label{squares}
For completeness and to facilitate the comparison with frustrated
clusters, we first summarize the main results obtained earlier for
small 2${\times}$2 and 2${\times}$4-sites bipartite clusters in
Refs.~\cite{JMMM,PRB,PLA,PRL}. The energies are measured in units
$|t|=1$ in all results that follow. Fig.~\ref{fig:gap-4-site}
illustrates $\Delta^c$ and $\Delta^s$ in ensemble of 2${\times}$2
square clusters at $\left\langle N\right\rangle \approx 3$ and
$T\to 0$. Vanishing of gaps indicates energy (multiple) level
crossings and corresponding quantum critical points, $U_c$ and
$U_F$. The negative gaps show phase separations for charge below
$U_c=4.584$ and spin degrees above
$U_F=18.583$~\cite{JMMM,Schumann1}. Phase A: Negative charge gap
below $U_c$ displays electron pairing $\Delta^P=|\Delta^c|$ and
charge phase separation into hole-rich (charged) metal and
hole-poor (neutral) cluster configurations. In a grand canonical
approach $\Delta^s>0$ at $U\leq U_c$ corresponds to electron
charge redistribution with opposite spin (singlet) pairing. This
picture for electron charge and spin gaps of equal amplitude
$\Delta^{s}\equiv\Delta^{P}=-\Delta^c$ of purely electronic nature
at $\left\langle N\right\rangle \approx 3$ is similar to the
BCS-like coherency in the attractive HM
and will be called coherent pairing (CP).
In equilibrium, the spin singlet background ($\chi_s>0$)
stabilizes phase separation of paired electron charge in a quantum
CP phase. The unique gap $\Delta^s \equiv\Delta^P$ at $T=0$ in
Fig.~\ref{fig:gap-4-site} is consistent with the existence of a
single quasiparticle energy gap in the BCS theory for
$U<0$~\cite{to be published}. Positive spin gap in
Fig.~\ref{fig:gap-4-site} at $U<U_c$ provides pairs rigidity in
response to a magnetic field and temperature (see
Sec.~\ref{diagram}). However, unlike in the BCS theory, the charge
gap differs from spin gap as temperature increases. This shows
that coherent thermal excitations in the exact solution are not
quasiparticle-like renormalized electrons, as in the BCS theory,
but collective paired charge and coupled opposite spins. The spin
gap $\Delta^s$ for excited $S={3\over 2}$ configuration in
Fig.~\ref{fig:gap-4-site} above $U_c$ is shown
for canonical energies 
in a stable MH-like state, $\Delta^c>0$. Phase B: Unsaturated
ferromagnetism (UF) for unpaired $S={1\over 2}$ with zero field
$\chi^s$ peak for gapless $s_z={\pm}{1\over 2}$ projections and
gapped $\Delta^{s}>0$ for $S={3\over 2}$ excitations at $U_c\leq
U\leq U_F$ will be called a spin liquid. Phase C: Negative
$\Delta^s<0$ at $U>U_F$ defines $S={3\over 2}$ saturated
ferromagnetism (SF). Localized holes at $\Delta^c>0$ rule out
possible Nagaoka FM in a metallic phase~\cite{Nagaoka}. Field
fluctuations lift $s_z$-degeneracy and lead to segregation of
clusters into magnetic domains.

It appears
 that the ensemble of square clusters share common and important
features with  larger bipartite clusters in the ground state and
at finite temperatures~\cite{to be published} (see
Sec.~\ref{diagram}).
For example, in 2x4 ladders  (Fig. 5 of Ref.~\cite{PRL}), we have identified
the existence of (negative and positive) oscillatory behavior in ($T=0$) charge gaps as
a function of $U$.
 Similar to what was seen in square clusters at low temperatures,
we
observe level crossing degeneracies in charge and spin
sectors  in bipartite 2${\times}$4 clusters at relatively
small and large $U$ values, respectively.
Thus the use of  chemical potential and departure from zero
temperature singularities in the canonical and grand canonical
ensembles appear to be essential for understanding important
physics related to the pseudogaps, phase separation, pairings and
corresponding crossover temperatures.
A full picture of coherent and incoherent pairing, electronic
inhomogeneities and magnetism emerges only at finite, but rather
low temperatures. (If we set $t=1$ eV, the most of the interesting
physics is seen to occur below a few hundred degrees K.)

\subsection{Tetrahedrons and square pyramids}
The topology of the tetrahedron is equivalent to that of a square with
next nearest neighbor coupling ($t^\prime=t$) while the square
pyramid of the octahedral structure in the HTSCs is related to
face centered squares (fcs). The average electron number
$\left\langle N\right\rangle$ and magnetization $\left\langle
s^z\right\rangle $ versus $\mu$ in Fig.~\ref{fig:num_mag} for
$T=0.01$ shows contrasting behavior in pairing and magnetism at
$t=1$ and $t=-1$ for the tetrahedron at $\left\langle
N\right\rangle=3$ and fcs at $\left\langle N\right\rangle=4$.
Different signs of $t$ in these topologies for one hole off half
filling lead to dramatic changes in the electronic structure.
Fig.~\ref{fig:gap} illustrates the
charge gaps at
small and moderate $U$. Tetrahedron at $t=-1$: SF with a negative
spin gap in a MH-like phase exists for all $U$. Tetrahedron at
$t=1$: metallic CP phase with charge and spin gaps of equal
amplitudes similar to the BCS-like pairing, discussed for squares
in Sec.~\ref{squares}, forms at all $U$. FCS at $t=-1$: MH-like
insulator displays two consecutive crossovers at $U \simeq 6.89$
from ($S=0$) antiferromagnetism (AF) into ($S=1$) UF and into
($S=2$) SF above $U = 12.19$. FCS at $t=1$: MH-like insulator
shows crossover at $U \simeq 29.85$ from ($S=0$) AF into ($S=1$)
UF. In triangles, SF and AF are found to be stable for all $U>0$
at $t=-1$ and $t=1$ respectively. Finally Table~\ref{Nagaoka}
illustrates magnetic phases
at large $U$ and $T=0$. For example, the squares and all
frustrated clusters at $t=-1$ exhibit stable SF; Tetrahedron and
triangle at $t=1$ retain CP and AF respectively; UF for the $S=1$
state, separated by $\Delta^s=-0.115$ from $S=0$, exists in fcs at
$t=1$. Fig.~\ref{fig:pyramid} shows charge gap at two coupling
values $c$ between the vertex and base atoms in the deformed
tetrahedron. Vanishing of
gap, driven by $c$, manifests
level crossings for $c=0.96$,
while $c=1.1$
and $t=1$ cases describe a single phase with avoided crossings.

\subsection{Phase diagrams}~\label{diagram}
Fig.~\ref{fig:ph4_6} illustrates a number of nanophases, defined
in Refs.~\cite{PRB,PRL}, for the tetrahedron similar to bipartite
clusters. The curve $\mu_+(T)$ below $T_c^P$ signifies the onset
of charge paired condensation. As temperature is lowered below
$T^*$, a spin pseudogap is opened up first, as seen in NMR
experiments~\cite{PRL}, followed by
the gradual disappearance of the spin excitations, consistent with
the suppression of low-energy excitations in the HTSCs probed by
STM~\cite{Bodi,hashini,Yazdani}. The opposite spin CP phase with
fully gapped collective excitations begins to form at $T\leq
{T_s}^P$. The charge inhomogeneities~\cite{Davis1,Valla} of
hole-rich and charge neutral {\it spinodal} regions between
$\mu_+$ and $\mu_-$ are similar to those found in the squares and
resemble important features seen in the HTSCs.
Fig.~\ref{fig:ph4_6} shows the presence of
bosonic modes below $\mu_+(T)$ and $T_s^P$
for paired electron charge and opposite spin respectively. This
picture suggests condensation of
electron charge and spin at various crossover temperatures while
condensation in
the
 BCS theory occurs at
 a single T$_c$ value. The temperature driven
spin-charge separation above $T_{s}^P$ resembles an incoherent
pairing (IP) phase seen in the
HTSCs~\cite{Bodi,Valla,hashini,Yazdani}. The charged pairs without
spin rigidity above $T_{s}^P$, instead of becoming
superconducting, coexist in a nonuniform, charge degenerate IP
state similar to a ferroelectric phase~\cite{to be published}. The
unpaired weak moment, induced by a field above $T_s^P$, agrees
with the observation of competing dormant magnetic states in the
HTSCs~\cite{hashini}. The coinciding $\chi^s$ and $\chi^c$ peaks
at $ T_{s}^P\leq T\leq T^\prime$ show full reconciliation of
charge and spin degrees seen in the HTSCs above T$_c$. In the
absence of electron-hole symmetry, the tetrahedral clusters near
optimal doping $\mu_P$ undergo a transition with
temperature from
a CP phase into a MH-like phase.
\bigskip

\section{Conclusion}
It is clear that our exact results, discussed above, provide novel
insight into level crossings, spin-charge separation,
reconciliation and full bose condensation~\cite{Lee}. Separate
condensation of electron charge and spin degrees offers a new
route for superconductivity, different from the BCS scenario. The
electronic instabilities found for various geometries, in a wide
range of $U$ and temperatures, will be useful for the prediction
of electron pairing, ferroelectricity~\cite{to be published} and
possible superconductivity in nanoparticles, doped cuprates, etc.
~\cite{Davis1,Valla,Bodi,hashini,Yazdani,deHeer0}. In contrast to
the squares, exact solution for the tetrahedron exhibits coherent
and incoherent pairings for all $U$. Our findings at small,
moderate and large $U$ carry a wealth of new information at finite
temperatures in bipartite and frustrated nanostructures regarding
phase separation, ferromagnetism and Nagaoka instabilities in
manganites/CMR materials. These exact calculations illustrate
important clues and exciting opportunities that could be utilized
when synthesizing potentially high-Tc superconducting and magnetic
nanoclusters assembled in two and three dimensional
geometries~\cite{TM}. Ultra-cold fermionic atoms in an optical
lattice~\cite{Fermi} may also offer unprecedented opportunities to
test these predictions.

We thank Daniil Khomskii, Valery Pokrovsky for helpful discussions
and Tun Wang for valuable contributions. This research was
supported in part by U.S. Department of Energy under Contract No.
DE-AC02-98CH10886.

\end{document}